\documentclass[aps,prb,twocolumn,superscriptaddress]{revtex4-1}

\usepackage{slashed}
\usepackage{appendix}
\usepackage{tikz-cd}
	\usepackage{amsmath}
	\usepackage{amsfonts}
	\usepackage{amssymb}
	\usepackage{graphicx}
	\usepackage{mathtools}
	\usepackage{stmaryrd}
	\usepackage{bm}
	\usepackage[colorlinks=true,citecolor=blue,linkcolor=red]{hyperref}
	\usepackage{bbold}					
	\usepackage[makeroom]{cancel}		
	\usepackage{multirow}				
	\usepackage[normalem]{ulem}        
	\usepackage{array}

\usepackage{soul}

	\newcolumntype{x}[1]{>{\centering\let\newline\\\arraybackslash\hspace{0pt}}p{#1}}
	
	\makeatletter
\newcommand*\rel@kern[1]{\kern#1\dimexpr\macc@kerna}
\newcommand*\widebar[1]{%
  \begingroup
  \def\mathaccent##1##2{%
    \rel@kern{0.8}%
    \overline{\rel@kern{-0.8}\macc@nucleus\rel@kern{0.2}}%
    \rel@kern{-0.2}%
  }%
  \macc@depth\@ne
  \let\math@bgroup\@empty \let\math@egroup\macc@set@skewchar
  \mathsurround\z@ \frozen@everymath{\mathgroup\macc@group\relax}%
  \macc@set@skewchar\relax
  \let\mathaccentV\macc@nested@a
  \macc@nested@a\relax111{#1}%
  \endgroup
}
\makeatother

	\DeclareMathAlphabet{\mathbbold}{U}{bbold}{m}{n}

	\def\bra#1{\left<{#1}\right|}				
	\def\ket#1{\left|{#1}\right>}				

	\newcounter{subeqn} %
	\makeatletter
	\@addtoreset{subeqn}{equation}
	\makeatother

\definecolor{XQ}{rgb}{1,0,0}
\definecolor{CW}{rgb}{.5,0,.5}
\definecolor{VS}{rgb}{0,0,1}
\definecolor{NA}{rgb}{0,1,0}

\begin{document}
\title{Signatures of Majorana bound states and parity effects in two-dimensional chiral $p$-wave Josephson junctions}

\date{\today}
\author{Nick Abboud}
\email{nka2@illinois.edu}
\author{Varsha Subramanyan} 
\author{Xiao-Qi Sun}
\author{Guang Yue}
\author{Dale Van Harlingen}
\author{Smitha Vishveshwara}
\email{smivish@illinois.edu}

\affiliation{Department of Physics, University of Illinois at Urbana-Champaign, Urbana, IL USA}

\begin{abstract}
We characterize topological features of Josephson junctions formed by coupled mesoscopic chiral p-wave superconducting islands. Through analytic and numerical studies of the low-lying BdG (Bogoliubov-deGennes) spectrum, we identify localized MBS (Majorana bound states) nucleated in Josephson vortices by the application of a perpendicular magnetic field. Additionally, we demonstrate the existence of an extended MBS that is delocalized around the outer perimeter of the coupled islands, which has measurable consequences on the Josephson supercurrent and phase dynamics of the junction. In particular, we predict a change in the critical current diffraction pattern in which the odd integer-flux nodes are lifted in a fermion parity-dependent fashion. We model the competing stochastic effects of thermal noise and macroscopic quantum tunneling within the RCSJ framework and show the emergence of a bimodal critical current distribution. We demonstrate that increasing the parity transition rate suppresses the bimodal nature of the distribution, thus strongly emphasizing the non-trivial parity dependent nature of the many-body ground state. Finally, we consider a trijunction geometry with three islands and discuss possible schemes to braid Majorana bound states by moving the Josephson vortices to which they are bound.

\end{abstract}

\maketitle

\section{Introduction} 

One of the most fascinating possibilities offered by two-dimensional topological systems is the existence of exotic fractionalized excitations--anyons--that are neither fermionic nor bosonic in nature. These excitations may be classified as Abelian or non-Abelian anyons based on their exchange statistics and braiding characteristics. Beyond their fundamental value, non-Abelian anyons are of interest as potential building blocks for robust, fault tolerant computation\cite{KITAEV,Freedman2002,Wang2001,Read}. Breakthroughs in the study of materials that can host anyons pave the path towards topologically protected quantum computing. The search for such platforms has emerged as an exciting nexus between condensed matter physics, material sciences, engineering, mathematics, as well as information sciences, promising to usher in an exciting new era of technology and processing capabilities\cite{Nayak}.
\begin{figure}[!h]
    \centering
    \includegraphics[width=0.48\textwidth]{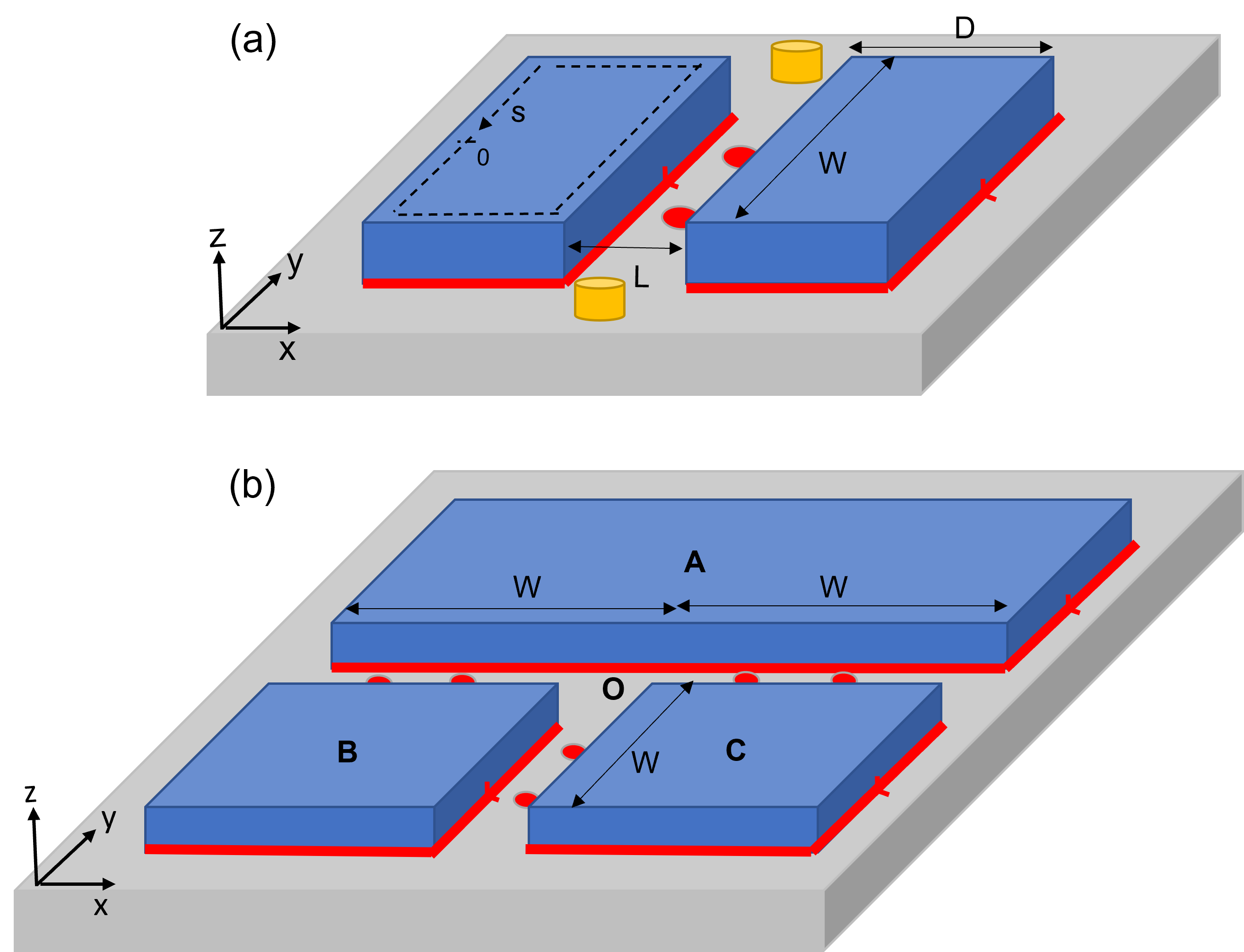}
    \caption{Lateral Josephson junctions composed of $p_x+ip_y$ topological superconductors (blue slabs) separated by a trivial conductor (grey slab) in a) two- and b) three-island geometries. Spontaneous chiral sub-gap Majorana modes (red lines) nucleate around the edges of the islands, perpendicular to the z-axis. Applying magnetic flux between the islands results in the formation of localized Majorana bound states (red circles).  a) Quantum dots (yellow blocks) coupled to the junction can act as detectors of electronic states shared by Majorana excitation pairs. b) The Majorana bound states may be adiabatically exchanged to perform a non-Abelian braiding operations, \textit{e.g.} by sequentially applying voltage pulses.}
    \label{2DJ}
\end{figure}

In recent times, Majorana bound states (MBS) have become the leading contenders for the formulation of topological qubits\cite{Alicea_2012,Stanescu_2013,Sarma2015}. Rapid developments in topological materials, capabilities for growing heterostructures, and nanowire technology, have resulted in a plethora of designs and proposals for nucleating and manipulating these low-lying excitations\cite{Lutchyn2010,Oreg2010,Alicea2011, Alicea_2012}. Superconducting nanowires have received much attention over the last decade as prospective candidates. Given the current obstacles in the nanowire system and the general need for having a selection to choose from, as with standard qubits,  it has become increasingly clear that multiple other platforms that can host MBS also need to be explored\cite{Sau, Alicea2016, Flensberg}. Presently these platforms, including the original $\nu=\frac{5}{2}$ states in fractional quantum Hall system revisited, as well as  lateral Josephson junctions, vortices in iron-based superconductors, and topological Floquet systems, are being extensively explored in both theory and experiments \cite{Lutchyn2010,Bravyi,Meyer2015,Stern2019,Dutta,Demler,Wang,Zhang,Zhang2019}. Ascertaining the existence of localized, controllable MBS and their braiding statistics in any system would make for a landmark achievement, not only from the quantum information perspective, but fundamentally in and of itself.

In particular, lateral Josephson junctions offer a promising alternative to their nanowire counterparts for localizing MBS even at low magnetic fields\cite{FuKane,FuPotter, Williams,Ye, Flensberg}. To summarize the key ideas, starting with the proposed setting by Fu and Kane\cite{FuKane}, the principle is to juxtapose appropriate thin film materials in patterned geometries so as to create networks of lateral topological Josephson junctions.  These junctions define proximity-coupled channels in which a field-induced phase winding can nucleate spatially separated Josephson vortices that host MBS. Pairs of MBS would together host an electronic state that could be occupied or unoccupied--the parity qubit.  The magnetic field can be adjusted to control the spacing of the MBS, and applied currents can be used to manipulate the positions of the MBS by controlling the phase degree of freedom between the two superconductors. Such a setup could be realized in a range of systems either composed of intrinsic topological superconductors or of superconducting gapped materials in contact with those possessing topologically protected gap\cite{Bernevig2013}. Thus, multiple material platforms and heterostructure configurations have been proposed as candidates, including proximity-induced superconductivity on 2D and 3D topological insulators, quantum anomalous Hall systems, quantum spin Hall systems, semiconductor-superconductor interfaces and so on\cite{FuKane,FuPotter,FuKane2009,Lutchyn2010,Stern2019,Meyer2015,Grosfeld2011,Pientka,Fornieri2019,Hegde2020, Flensberg}. 

As with established nanowire geometries, a complete architecture has been recently proposed to manipulate the MBS within a lateral Josephson junction and perform gate operations including by three authors of this article\cite{Hegde2020,Pientka}. Previous work considers an extended junction formed by S-TI-S heterostructures, which continue to be extensively studied and characterized via experiments\cite{Kurter, Stehno}. Here, we turn to the essential task of laying out and analyzing crucial steps leading to such a functioning device. While several of these components have been investigated in different contexts, our purpose here is to have a focused treatment in the specific context of lateral junctions, as directly relevant to current experimental capabilities, including on-going work \cite{Stehno, Sochnikov, Williams,Ye,Kayyalha,Kurter2015,sedlmayr2020dirac}. We center our analysis on a simple, minimal model of a lateral topological Josephson junction. We consider spinless $p_x+ip_y$ superconducting islands separated by a trivial insulator \cite{Grosfeld2011,Alicea_2012,Grosfeld}. We address two key questions - (a) In the presence of a perpendicular magnetic field, where are the MBS precipitated in such systems and what are the effects of finite size? (b) What are some key signatures of the presence of MBS and associated parity states, and how can they be experimentally accessed? Additionally, we propose parity readout schemes and braiding protocols that may be easily adapted across platforms. As with nanowire geometries, where the very first issue of the zero-bias conductance peak associated with the Majorana-based midgap electronic state is itself in question, experimentally ascertaining any of these steps would constitute a leap in progress. 

Our approach involves considering the lateral junction in two geometries, as shown in Fig.\ref{2DJ} - the two island geometry, with a single junction formed by a trivial conductor (or trivial insulator with a conducting surface state) between the topological superconductors, and the three island geometry, analogous to the nanowire Y junctions \cite{Alicea_2012}. In the geometry composed of two mesoscopic superconductor islands, we chart out the nucleation of Majorana bound states as a function of system parameters, and show delocalization along the islands' shared outer perimeter. It must be pointed out that a similar result was obtained for S-TI-S junctions in the presence of a perpendicular magnetic field in \onlinecite{choi2018,choi2019}. Here, we show the same for $p_x+ip_y$ junctions and further propose detection via coupling to quantum dots, a step akin to measuring zero-bias conductance peaks in nanowires.  In diffraction patterns related to critical current as a function of applied flux, we demonstrate an electron parity-dependent node lifting as well as a bimodal switching current distribution across the junction, reminiscent of the fractional Josephson effect, which requires quantum coherence and charge $e$ processes \cite{VonOppen}. We consider three different stochastic processes that could affect the current and identify regimes in which the bimodal signature would remain robust. Finally, as the smoking gun signature of Majorana bound states, we adapt the nanowire T-junction scheme to the nanowire setting as a means to perform exchanges that result in non-Abelian braiding. 

In what follows, we present the model Hamiltonian for describing the mesoscopic topological superconducting island setup in Section \ref{Models}. We provide a discretized version amenable for numerical treatment as well as an effective one-dimensional model that focuses on the low-lying dispersing modes along the periphery of the islands which are responsible for the MBS. For the two-island setting, we obtain the spectrum of states and pinpoint the modes corresponding to the MBS as a function of applied flux in Section \ref{locating}. We then employ this information for obtaining the parity-dependent critical Josephson current across the junction in Section \ref{parity}. To model the fate of this current and its bimodal nature in the presence of geometric capacitance and resistive quasiparticle flow, we employ the standard RCSJ model, adapted here to account for parity. We then turn to the three-island geometry in Section \ref{trij} to outline the protocols required for braiding. 

\section{Models}\label{Models}
The geometries we consider are all modeled within the framework of a spinless chiral topological superconductor\cite{Stone2004, Bernevig2013} described by the Nambu-space Bogoliubov-de Gennes (BdG) Hamiltonian

\begin{align}
	\mathcal H = \begin{pmatrix} -\frac{1}{2m^*}\bm \nabla^2 - \tilde \mu & \hat \Delta \\ \hat \Delta^\dag & \frac{1}{2m^*}\bm \nabla^2 + \tilde \mu  \end{pmatrix}. \label{hbdg}
\end{align}
The spatially varying electrochemical potential is given by ${\tilde \mu(\bm r) = \mu(\bm r) - V(\bm r)}$, where $V(\bm r)$ denotes a local external potential. A representative chiral ${p_x+ip_y}$ pairing operator is taken to have the form\cite{Stone2004} ${\hat \Delta = i\frac{\Delta(\bm r)}{k_F} e^{i\varphi(\bm r)/2}(-i\partial_x+\partial_y)e^{i\varphi(\bm r)/2}}$, where  $\varphi(\bm r)$ denotes the superconducting phase.
We also define a velocity by $\hbar v(\bm r) = \frac{\Delta(\bm r)}{k_F}$. The topological and trivial phases of this model in the bulk are distinguished by the value of $\tilde\mu$; the topological phase has $\tilde\mu > 0$ while the trivial phase has $\tilde\mu < 0$. By appropriately choosing the parameters' spatial variation in the plane, we can construct different device geometries.

We first study a Josephson junction of width $L$ formed by bringing together two mesoscopic $p_x+ip_y$ superconducting islands separated by a trivial insulator, as shown in Fig. \ref{2DJ}a. The function $\tilde\mu(\bm r)$ is set to a positive constant $\tilde\mu_0$ within the islands and to $-\tilde\mu_0$ elsewhere; likewise, $\Delta(\bm r)$ is equal to $\Delta_0$ within the islands and vanishes elsewhere.

We assume that a magnetic field $B$, applied perpendicular to the plane, is completely expelled from the islands (zero London penetration depth). At the same time, the self-field of the Josephson currents is assumed to be sufficiently small to prevent the expulsion of field from the junction region separating the islands; this is known as the short junction limit\cite{tinkham2004}. The short junction limit can be stated as $J_c \ll \Phi_0/2\pi\mu_0LW^2$, where $\Phi_0=h/2e$ and the Josephson current density $J_c$ depends on the extent to which the supercurrent is spread out in the $z$-direction. We expect that a window of supercurrent values would respect both this limit as well as the quasi-2D geometry. 

The total flux within the junction is uniform and takes the value $\Phi=BLW$. In a Landau gauge, the corresponding vector potential is $A_x=0$ and
\begin{align}
    A_y &= \begin{cases}
                -\frac{\Phi}{2W} & x \le 0 \\
                -\frac{\Phi}{2W} + \frac{\Phi}{W}\frac{x}{L} & 0 < x < L \\
                \frac{\Phi}{2W} & x \ge L
           \end{cases}.
\end{align}
The ansatz for the superconducting phase $\varphi(\bm r)$ is chosen such that the Landau-Ginsburg supercurrent density ${\bm j_s \propto \bm \nabla \varphi - \frac{2\pi}{\Phi_0}\bm A}$ vanishes within the islands. This gives
\begin{align}
    \varphi(\bm r) &= \begin{cases}
                        -(2\pi\frac{\Phi}{\Phi_0}\frac{y}{W} + \phi_0)/2 & x \le 0 \\
                        (2\pi\frac{\Phi}{\Phi_0}\frac{y}{W} + \phi_0)/2 & x \ge L
                    \end{cases}, \label{phiofr}
\end{align}
where $\phi_0$ is a parameter describing the inter-island superconducting phase difference at $y=0$.

Some of our treatment below involves full numerical diagonalization in this two-island geometry. We employ a discretized version of the Hamiltonian on a square lattice with spacing $a$,
\begin{align} \label{H2D-discrete}
    \mathcal H_D &= \sum_{i, j}\biggl[ (4t-\tilde\mu_{i, j})\ket{i, j}\bra{i, j}\tau_z + \\ \nonumber
    &\biggl( \ket{i+1, j}\bra{i, j}\mathbb H_x + \ket{i, j+1}\bra{i, j} \mathbb H_y + \text{H.c.}\biggr) \biggr],
\end{align}
where
\begin{align}
    \mathbb H_{x/y} &= -te^{i\tau_z\theta^{x/y}_{i, j}}\tau_z - i\frac{\hbar v_{i,j}}{2a}e^{i\tau_z\varphi^{x/y}_{i,j}}\tau_y
\end{align}
Here $\tau_x$, $\tau_y$ and $\tau_z$ are Pauli matrices acting in the particle-hole space, and the hopping parameter takes the standard form $t = \frac{\hbar^2}{2m^*a}$. The superconducting phases $\varphi_{i,j}^x$ and $\varphi_{i, j}^y$ are defined on the links. Inclusion of the Peierls phases $\theta_{i,j}^x$ and $\theta^y_{i,j}$ does not make a qualitative difference to the structure of the midgap modes of interest. 

As has been well studied\cite{Alicea_2012}, the sub-gap degrees of freedom in the junction geometry are chiral Majorana fermions propagating along the one-dimensional edges of the two islands. To obtain a heuristic understanding of the sub-gap spectrum and wavefunctions, we discard the kinetic energy term in Eq.\ \ref{hbdg} and project the Hamiltonian to the space spanned by the chiral Majoranas. 

One may use a common coordinate system, denoted here by $s$, to obtain the peripheral Hamiltonian\cite{Grosfeld2011}
\begin{align}
	\hat H_\text{eff} &= \frac{1}{2}\int ds\ \begin{pmatrix} \psi_L & \psi_R \end{pmatrix} \underbrace{\begin{pmatrix} iv \partial_s && -iW(s) \\ iW(s) && -iv \partial_s \end{pmatrix}}_{\equiv \mathcal H_\text{eff}} \begin{pmatrix} \psi_L \\ \psi_R \end{pmatrix}.\label{heff}
\end{align}
Here $\psi_{L}(s)$ and $\psi_R(s)$ are Hermitian fields describing the chiral Majorana fermions confined to the edges of the left and right islands, respectively. The term ${W(s) = m(s)\cos\bigl(\frac{\phi(s)}{2}\bigr)}$ mixes the counterpropagating $\psi_{L}$ and $\psi_R$ in the junction region, and  $\phi(s)=\varphi_R(s)-\varphi_L(s)$ is the (gauge-invariant) local superconducting phase difference across the junction at position $s$. If one of the islands is made of $p_x-ip_y$ paired superconductor, the chiral modes co-propagate along the edges of the islands. The mixing term then takes the form ${W(s) = m(s)\sin\bigl(\frac{\phi(s)}{2}\bigr)}$\cite{Grosfeld2011}. Here, we focus on the system where both islands are composed of $p_x+ip_y$ paired superconductors. From Eq.\ \ref{phiofr},
\begin{align}
    \phi(s) = 2\pi\frac{\Phi}{\Phi_0}\frac{s-P/2}{W} +\phi_0. \label{phiofs}
\end{align}
The tunneling function $m(s)$ vanishes outside the junction region; for simplicity we take $m(s) = m_0\Theta\bigl(\frac{W}{2}-\bigl|s-\frac{P}{2} \bigr|\bigr)$,
where $P= 2(W+D)$ is the perimeter of each island. The tunneling parameter $m_0$ decreases exponentially with increasing $L$. Crucial to what follows, we assume that no Abrikosov vortices penetrate the bulk of either island. Consequently, the boundary conditions appropriate to $\psi_L$ and $\psi_R$ are antiperiodic: $\psi_L(s+P) = -\psi_L$ and $\psi_R(s+P) = -\psi_R$.

\begin{figure}[!h]
    \centering
    \includegraphics[width=.48\textwidth]{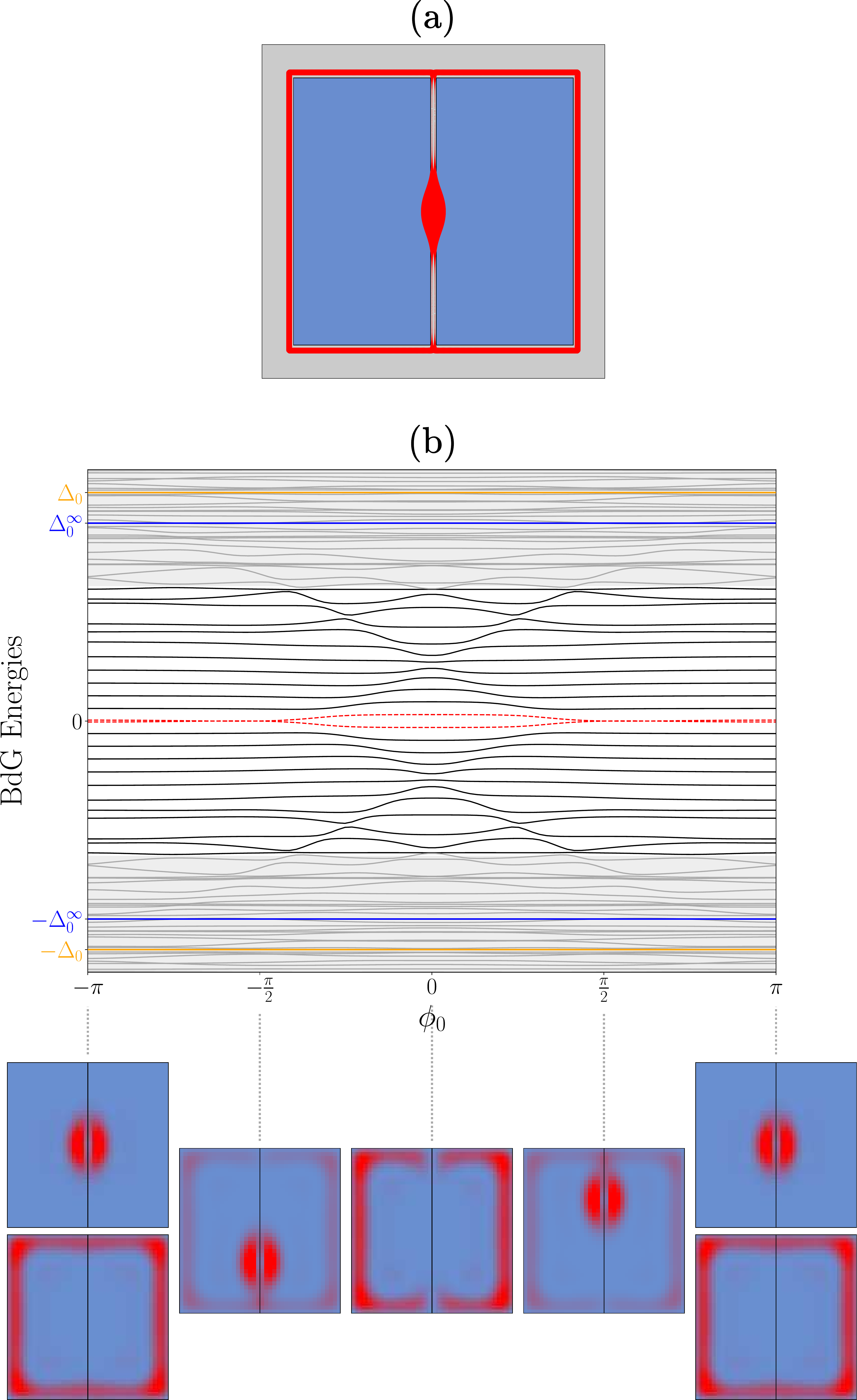}
    
    \caption{\small Spectrum and wavefunctions. (a) Weight of the lowest quasiparticle wavefunction (red) obtained by simulation of the one-dimensional effective model $\hat H_\text{eff}$ when $\Phi/\Phi_0=1/2$ and $\phi_0=\pi$, exhibiting a MBS localized within the junction and a MBS delocalized around the islands' shared perimeter. The line width is proportional to the square of the wavefunction weight. (Numerical parameters: $W=40a$, $D=20a$, $L=a$, $m_0=10$, $v=1/30$, Energy scale set by $\Delta_0$). (b) Numerical BdG spectrum as a function of $\phi_0$ when $\Phi=1.3\Phi_0$. (Numerical parameters: $W=1\ \mu\text{m}$, $D=0.5\ \mu{m}$, $L=a=0.025\ \mu\text{m}$, $\Delta_0=\tilde\mu = 100\ \mu\text{eV}$, $m^*=0.15m_e$) W= States in the shaded regions extend into the bulk of the islands, while those in the unshaded region are localized within the junction and on the edges. For the chosen parameters, $\Delta_0^\infty=\frac{\sqrt{3}}{2}\Delta_0$ is the islands' bulk gap in the infinite-area limit. Weights of lowest quasiparticle wavefunctions (red), corresponding to the red dotted BdG levels, are shown superimposed on the junction at various values of $\phi_0$. As $\phi_0$ increases from $-\pi$ to $0$, a localized/delocalized pair of MBS fuses into a non-zero-energy complex-fermion quasiparticle.}
    \label{fig:spectrum}
\end{figure}

\section{Locating the Majoranas}\label{locating}
The effective one-dimensional model serves well to provide an immediate characterization of the MBS configurations. In the presence of magnetic flux, a single MBS is present in each Josephson vortex, that is, at each point in the junction where the phase difference $\phi(s)$ equals an odd multiple of $\pi$. By inspection of Eq. \ref{phiofs}, these points are equally spaced along the junction at $s_n = \frac{P}{2} + \bigl(n+\frac{1}{2} - \frac{\phi_0}{2\pi}\bigr)l_B$, where $l_B = \frac{W}{\Phi/\Phi_0}$ is the magnetic length separating adjacent Josephson vortices.

In the vicinity of the Josephson vortex at $s_n$, to obtain the form of an isolated MBS, one can linearize the coupling $W(s)$ in Eq. \ref{heff} and solve for the zero mode operator, $[\gamma_n, \hat H_\text{eff}] = 0$. The Majorana bound state thus obtained is 
\begin{align}
    \gamma_n = \int ds\ \frac{\exp\bigl[ - \frac{1}{2}\bigl(\frac{s-s_n}{\lambda_M}\bigr)^2\bigr]}{(8\pi\lambda_0^2)^{1/4}}(\psi_R +(-1)^n \psi_L), 
\end{align}
which satisfies the Hermitian requirement $\gamma_n^\dag=\gamma_n$. The linearization is valid for this purpose when the localization length $\lambda_M = \sqrt\frac{Wv}{\pi(\Phi/\Phi_0)m_0}$ is much smaller than the magnetic length $l_B$. The total number of MBS is proportional to the amount of flux penetrating the junction, while the absolute position of the equally-spaced array of MBS depends on $\phi_0$. The MBS can thus be moved rigidly along the junction by varying $\phi_0$, e.g. by means of applied voltage pulses\cite{Hegde2020} or by connecting the two superconductors through a SQUID-based loop and passing extra flux.

In the context of the finite-size geometries we study, an interesting feature emerges due to the requirement that MBS occur in pairs. Whenever there is an odd number of MBS localized within Josephson vortices, one may ask where an extra zero mode is located. Similarly to the case of bound states vortex cores within a single superconducting island \cite{Bernevig2013,Grosfeld2011,Ivanov,Stone2004}, the answer lies in the periphery. Sub-gap quasiparticles delocalized around the device's shared outer perimeter form a near continuum in which the energetic level spacing is inversely proportional to the linear size of the device. When the junction contains an odd number of Josephson vortex-bound MBS and no Abrikosov vortices pierce the islands themselves, one finds that that this near continuum contains an additional Majorana zero mode. This behavior may be established analytically\cite{choi2019}. We remark that the strictly 2D setup necessitates such a scenerio; a system of finite thickness could admit a vortex line permeating through the third dimension, harboring pairs of MBS at the two ends\cite{ching-kai2011}.

To demonstrate these features explicitly, we have carried out numerical diagonalization of the effective 1D junction Hamiltonian in Eq. \ref{heff}. Care must be taken to subtract the doubled chiral Majorana fermions that appear at momentum $k=\pi/a$ when naively discretizing Eq.\ \ref{heff} on a lattice with spacing $a$. We elected to Fourier transform Eq.\ \ref{heff} on a ring and then impose a momentum cutoff $|k|< \Lambda$, where $\Lambda \gg \lambda_M^{-1}$ to ensure the ability to spatially resolve the low-energy modes. A sample low-lying mode thus obtained, which can be interpreted as a hybridized state between a MBS localized within a Josephson vortex and a MBS delocalized around the periphery, is shown in Fig.\ \ref{fig:spectrum}a. On continuously tuning $\phi_0$ so that the array of an odd number of vortex-localized MBS traverses along the junction, we observe a hybridization event whenever a vortex-localized MBS approaches the MBS delocalized around the junction. During the course of this hybridization, the localized/delocalized MBS pair is transformed into a single complex fermion having a non-zero energy inversely proportional to the joint outer perimeter of the two islands.

The qualitative features of the low-lying sub-gap states obtained from the the 1D effective description Eq.\ \ref{heff} are also borne out by numerical diagonalization of the full 2D Hamiltonian of Eq. \ref{H2D-discrete}, as shown in Fig.\ \ref{fig:spectrum} for $\Phi/\Phi_0 \gtrsim 1$. If $\phi_0=-\pi$, then the lowest quasiparticle state has zero energy and is a very weakly hybridized state of two MBS--one vortex-localized and one delocalized along the perimeter. If $\phi_0$ is increased, the vortex-localized MBS migrates to the lower end of the junction, whereupon it hybridizes with the delocalized MBS to form the non-zero energy complex quasiparticle depicted when $\phi_0=0$. Further increasing $\phi_0$ causes the reverse process to occur at the upper end of the junction. Apart from the low-lying states bound to the junction and perimeter, bulk states with support in the islands are present at energies above the bulk gap. This 2D simulation forms the basis of our analysis of Josephson currents in the next section.

Turning to possible measurements these observations can lead to, there have been multiple proposals to detect the presence of MBSs in Josephson junctions in nanowires \cite{VonOppen}. Some proposals attempt to take advantage of the non-local parity of MBSs and its manifestation in a number of measurable physical phenomena \cite{Deng,Zocher,Zhang21,Gharavi}. Alongside these methods, electron parity may also be directly detected through quantum dots and single-electron transistors (SETs). In this lateral junction situation, we remark that signatures in Josephson currents and single-electron detection are the most basic indicators of the existence of potential MBSs, similar to nanowire zero-bias conductance peaks--a necessary but not sufficient condition. The geometry here also provides a unique setup for observing the interplay between the MBSs and also with the delocalized Majorana mode. We thus propose positioning quantum dots as shown in Fig. \ref{2DJ} as well as over the junction above the plane of the islands. Voltage pulses can induce MBS pairs to be driven towards the latter kind of quantum dots. The parity of the complex fermion state formed by a MBS pair (occupied or unoccupied) can be determined by measuring the conductance shift associated with the quantum dot\cite{Zhang19}. Likewise, the quantum dots shown in Fig. \ref{2DJ} can register the parity-based hybridization of an MBS with the delocalized mode.

\begin{figure*}
\centering
    \includegraphics[width=\textwidth]{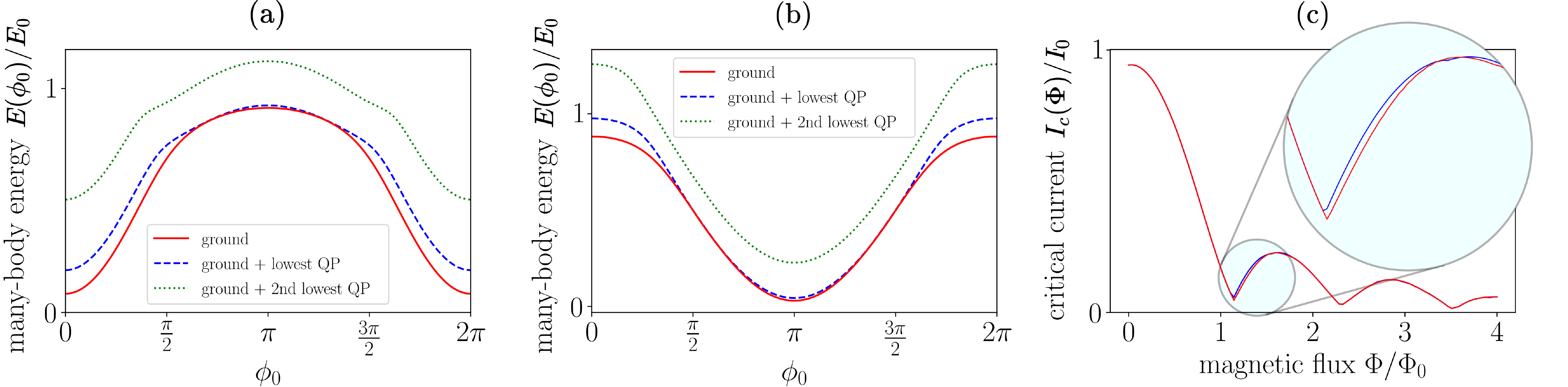}
    \caption{ Many-body energy of the junction in its $\phi_0$-dependent instantaneous ground state (solid red), lowest quasiparticle occupation above the filled ground state (dashed blue) and similarly occupation of the second-lowest quasiparticle (dotted green) when (a) $\Phi=\Phi_0$ and (b) $\Phi=1.3\Phi_0$. In (a), low-temperature thermal equilibrium is achieved in a non-degenerate state at $\phi_0=0$; in (b), it is achieved in a nearly degenerate state at $\phi_0=\pi$. (c) The critical current diffraction pattern splits into parity-dependent branches near odd-integer flux and exhibits both parity-dependent and parity-independent node lifting. Device parameters chosen as in Fig. 2. The energy axis in (a) and (b) has been shifted and scaled by $E_0=30\ \mu\text{eV}$. The critical current in (c) has been scaled by $I_0=40\ \text{nA}$.}
    \label{WP}
\end{figure*}
\section{Parity signatures in the critical current distribution}\label{parity}
Another signature of Majorana fermion-based parity effects concerns the behavior of supercurrent flowing across the junction. In this section, we present a numerical study of Josephson current in the finite-sized two-island junction of the previous section. We focus on the critical current---the maximum current bias current at which the voltage remains zero--and its behaviour under various conditions. We predict MBS-associated bimodality in the critical current diffraction pattern corresponding to two different values of critical current that depend on the parity of the system. Having shown that there exist two possible values of critical current that depend on parity, more generally, we demonstrate that the bimodality can remain detectable even when considering the effects of dissipation, charging energy, finite temperature, macroscopic quantum tunneling and quasiparticle poisoning. We model the stochastic nature of the associated critical current distribution by taking into account temperature-activated escaping events as well as macroscopic quantum tunneling. Related universal behaviour has been studied both experimentally and numerically, particularly in the context of nanowires and graphene-based systems.\cite{Sahu2009,Finkelstein,Levchenko1,Levchenko2,Guarcello_2017,Murphy2015}. The bimodality studied here can serve as an indicator of the existence of MBS, although it is insensitive to their exchange statistics.
\subsection{Derivation of Critical Current from Energy Spectrum}
We first derive the critical current $I_c$ as a function of the applied flux from the energy spectrum derived in the previous section, as is standard for any given spectrum corresponding to a Josephson junction setting\cite{waldram2017, tinkham2004, kwon2004}
\begin{align} \label{ic}
    I_c=\max_{\phi_0}\biggl\{\frac{2e}{\hbar}\frac{\partial E}{\partial\phi_0}\biggr\},
\end{align}
where $E(\phi_0)$ is the many-body energy of the two-island device followed adiabatically as a function of $\phi_0$. In the $\phi_0$-dependent ground state,
\begin{align}
    E_\text{gnd}(\phi_0)=-\frac{1}{2}\sum_{\epsilon_n>0}\epsilon_n(\phi_0),
\end{align}
where $\epsilon_n(\phi_0)$ are the positive quasiparticle energies at $\phi_0$ obtained from the numerics of Fig. \ref{fig:spectrum}. The energies of higher states are obtained from $E_\text{gnd}$ by the addition of one or more of the $\epsilon_m>0$.

In accounting for finite but small temperature, the question arises as to which state(s) one should use to evaluate Eq.\ \ref{ic}. First, as the most physically relevant possibility, we propose that the appropriate state is determined by initializing the junction in equilibrium with a thermal/particle bath. In the regime of flux considered here, $\Phi/\Phi_0\approx 1$, we may focus on the two lowest-energy many-body states: the ground state with energy $E_\text{gnd}(\phi_0)$ and the first excited state with energy $E_\text{1}(\phi_0)$. These two states differ by the addition of the lowest-lying quasiparticle, which as we have seen is associated to the MBS. Correspondingly, the two states have opposite fermion number parity. In Fig.\ \ref{WP}ab, $E_\text{gnd}(\phi_0)$ and $E_1(\phi_0)$ are depicted by the solid red and dashed blue curves, respectively. We assume that the temperature $k_BT$ of the bath is large compared to the exponentially small hybridization energy of well-separated MBS (as in the $\phi_0=-\pi$ wavefunction of Fig.\ \ref{fig:spectrum}b), but small compared to the hybridization energy between the localized and delocalized MBS when they are brought together to overlap (as when $\phi_0=0$). For the specific set of parameters chosen in our simulations, these constraints can be met if $k_BT\ll\Delta_0$. If $\Phi/\Phi_0 \lesssim 1$ as in Fig.\ \ref{WP}a, then the junction equilibriates to the energetic minimum near $\phi_0\approx0$. Since $E_1(0)-E_\text{gnd}(0) \gg k_BT$, the equilibrium many-body state is well approximated by the pure ground state, ignoring the negligibly small mixture with higher states. Hence we use $E_\text{gnd}$ in Eq.\ \ref{ic} to compute $I_c$ at this flux. If $\Phi/\Phi_0 \gtrsim 1$ as in Fig.\ \ref{WP}b, then the junction equilibrates to $\phi_0\approx\pi$. Since $E_1(\pi)-E_\text{gnd}(\pi) \ll k_BT$, the equilibrium many-body state is well approximated by the completely incoherent, equiprobable mixture of the ground state and first excited state. Hence at this flux we obtain two values for $I_c$ with equal probability, computed by using $E_\text{gnd}$ or $E_1$ in Eq.\ \ref{ic}.

Proceeding in this manner to compute $I_c(\Phi)$, we obtain the critical current diffraction pattern of Fig.\ \ref{WP}c. Interestingly, the distribution  exhibits parity-dependent lifting of the odd integer nodes as well as parity-independent lifting of all nodes. At even nodes, in-junction localized MBSs simply trade places with the MBS delocalized around the perimeter as a function of phase difference $\phi_0$, and no hybridization takes place. For this reason, we expect no parity dependence of the critical current near even nodes, though node lifting is still present. The MBS hybridization at the junction ends only occurs when the value of the flux is near an odd node of the diffraction pattern, leading to the parity-dependent node lifting.

We mention two caveats to the results obtained in Fig.\ref{WP}. First, our intent is to show that the two values of parity-dependent critical current can indeed be discernible in principle even in the presence of the mid-gap states characteristic of extended junctions. However, not all parameter ranges show a clear splitting. Second, the nodes of the diffraction pattern are expected to be located at integer values of $\Phi/\Phi_0$ for large systems. In our numerics, we find the spacing between nodes to be slightly larger than one flux quantum, as visible in Fig.\ \ref{WP}c, which we expect to be due to the small size of our system. In fact, this effect is highly reduced in simulating physically larger geometries with the same number of lattice sites, but the parity dependence also reduces. This observation is indicative that parity dependence would require geometries employing mesoscopic grains. 

In comparing our work to that of the original setting proposed by Potter and Fu, Ref. \cite{FuPotter} predicts a similar MBS-attributed node lifting in a 3DTI/s-wave heterostructure. Apart from differences in the considered material and geometry, our analysis differs from theirs in that it treats the occupation of a MBS pair as a long-lived degree of freedom on the timescale of critical current measurement. We have found that this leads to additional parity dependence of the critical current diffraction pattern. Furthermore, the difference in geometry between our system and potter and Fu's leads to a differing prediction for the MBS-attributed critical current. In Potter and Fu's geometry, pairs of MBSs on the top and bottom of the device can hybridize around the ends of the junction no matter the value of the magnetic flux, whereas in our analysis this hybridization, and concomitant parity-dependent current contribution, only occurs near odd nodes.

To highlight the key result we study here, as exhibited in  Fig.\ \ref{WP},  the MBS parity degree of freedom changes the current-phase relation of the junction. This change is reflected in the magnitude and shape of the Josephson washboard potential and effectively a parity-dependent critical current. This dependence can be probed by measuring the distribution of the transition to the finite voltage state as the bias current is increased, manifesting in a bimodal distribution. Observing this bimodality would make for a prominent and direct indication of parity-dependent physics, which in this case is tied to the presence of MBS. 

\subsection{RCSJ modeling of stochastic processes}
As with conventional superconductors, more detailed and realistic predictions for critical current distributions can be generated by taking into account the geometric capacitance and resistive quasiparticle flow across the junction using the RCSJ circuit model\cite{tinkham2004, mccumber1968, stewart1968, ambegaokar1969}. In particular, we aim to address the robustness of the critical current bimodality in the presence of multiple physical processes. Similar studies in related systems have been carried out by Refs.~\onlinecite{peng2016, shu-ping2014}. To this end we adopt the RCSJ model, in which the ideal Josephson junction is placed in parallel with a lumped-element capacitor (capacitance $C$) and resistor (resistance $R$). We first provide a general analysis and expectations for bimodality, adhering to the experimentally viable parameters employed in our experience. We then demonstrate the ubiquitous nature of bimodality by considering another range of parameters, namely those that match the order of magnitude critical current parameters obtained from the energy spectrum in the subsection above. 

\begin{figure}
    \centering
    \includegraphics[width=.42\textwidth]{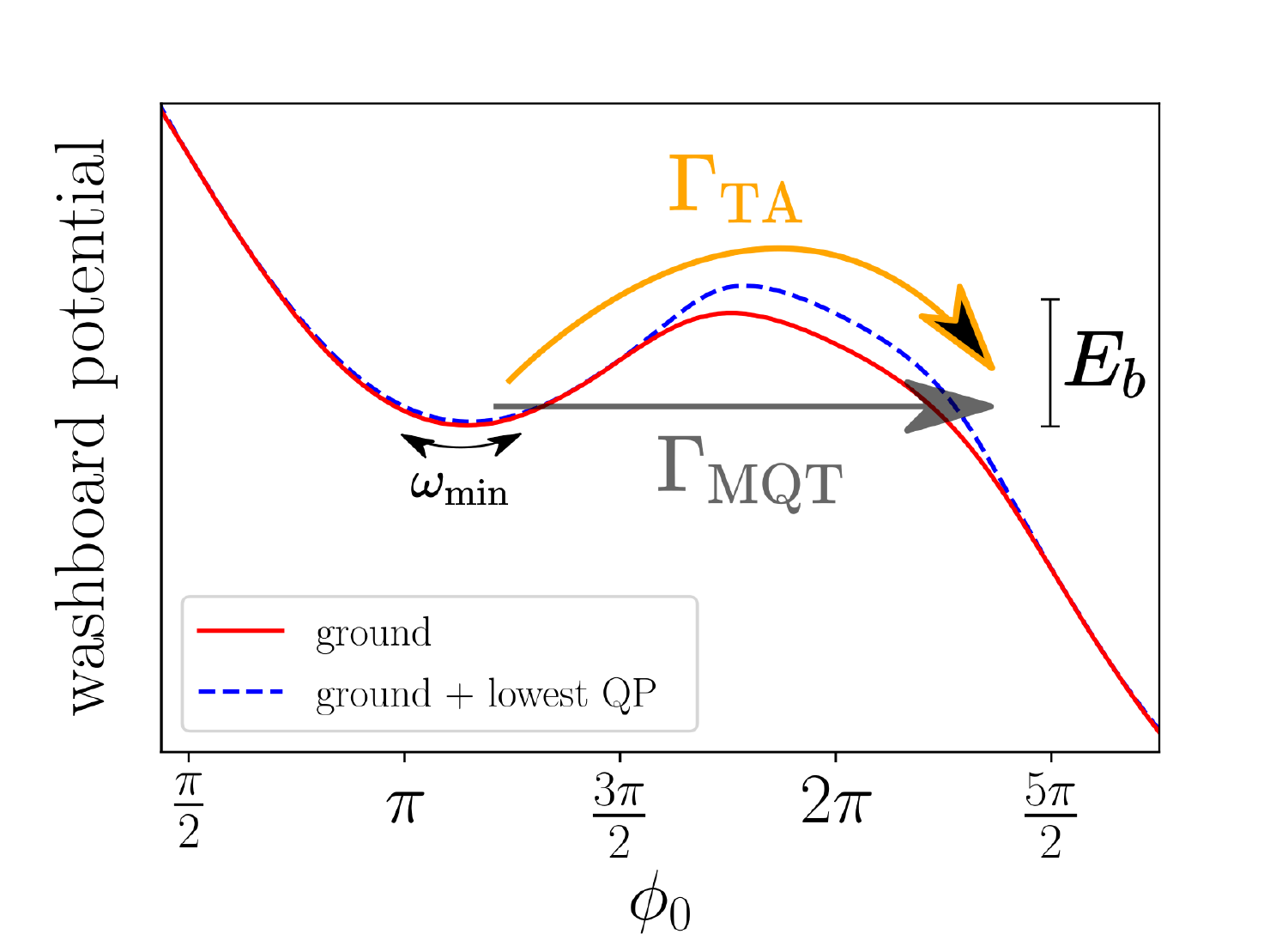}
    \caption{Depiction of the two tilted washboard potentials differentiated by their associated parities and various stochastic processes that could act on a state that is initially at a minimum of the potential in the RCSJ model. Such a state could undergo macroscopic quantum tunneling at rate $\Gamma_{MQT}$ or undergo thermally-activated escape over the energy barrier energy $E_b$ at rate $\Gamma_{TA}$. 
    \label{fig:RCSJ}}
\end{figure}

In the RCSJ model, the dynamical behavior of $\phi_0$ maps onto that of the position of a particle of mass $M=\bigl(\frac{\Phi_0}{2\pi}\bigr)^2C$ subject to a linear damping force having a characteristic timescale $\tau=RC$ and a one-dimensional washboard potential $U(\phi_0)=E(\phi_0)-\frac{\Phi_0}{2e}I_\text{bias}\phi_0$. The washboard potential can be tilted by biasing the junction with applied  current $I_\text{bias}$. Note that the washboard potential differs for the different many-body states (see Fig.\ \ref{fig:RCSJ}). The equation of motion describing this model is
\begin{align} \label{rcsj}
    M\ddot\phi_0 = -U'(\phi_0) - \frac{M}{\tau}\dot\phi_0.
\end{align}
Here, an overdot denotes differentiation with respect to time, and a prime denotes differentiation with respect to $\phi_0$. At time $t=0$, $I_\text{bias}=0$ and one assumes $\phi_0$ is trapped in a minimum of the washboard potential. Upon ramping $I_\text{bias}$, as it approaches the critical value $I_\text{c}$ the local minimum of $U(\phi_0)$ vanishes, $\phi_0$ escapes to a running state with nonzero $\dot\phi_0$, and the junction develops a finite voltage $V\propto \dot\phi_0$ according to the Josephson relation. This is the behavior expected from Eq.\ \ref{rcsj} in its overdamped regime in which $Q\approx\omega_0RC>1$, where $\omega_0$ denotes the oscillation frequency of the unbiased washboard potential. Here, the experimentally measurable critical current $I_c$ is obtained as the value of $I_\text{bias}$ at which the junction develops a finite voltage. 

The situation here warrants the further inclusion of three  stochastic effects. We introduce a rate $\Gamma_\text{TA}$ for thermally-activated escape from the washboard potential minimum and a rate $\Gamma_{MQT}$ for macroscopic quantum tunneling through the potential barrier. In the low-damping regime relevant to our devices, the thermal activation rate can be determined in a saddle-point approximation to have the Arrhenius form\cite{peng2016}
\begin{align}
    \Gamma_\text{TA} = \frac{\omega_p}{2\pi} e^{-E_b/k_BT_\text{eff}}.
\end{align}
Here $\frac{\omega_p}{2\pi}$ can be thought of as an attempt frequency, and $E_b$ is the barrier height for escape from the well.  These depend implicitly on $I_\text{bias}$. The temperature $T_\text{eff}$ parameterizing thermal noise can in principle exceed the device temperature $T$, e.g. if the external leads are at a higher temperature than the device itself. For macroscopic tunneling we use the expression derived by Caldeira and Leggett \cite{caldeira1983, devoret1985},
\begin{align}
    \Gamma_\text{MQT} = &\frac{\omega_p}{2\pi}\biggl[120\pi\biggl(\frac{7.2E_b}{\hbar \omega_p}\biggr)\biggr]^{1/2}\\
    &\quad\times \exp\biggl[-\frac{7.2E_b}{\hbar \omega_p}\biggl(1+\frac{0.87 }{\tau\omega_p}\biggr)\biggr].
\end{align}

The third effect is the parity transition rate $\Gamma_\text{P}$ characterizing changes in the parity state of Majorana pairs.  This could arise from effects such as quasiparticle poisoning of the low-lying modes that can trigger a transition between the washboard potentials of the ground and first excited many-body states.

To generate the probability distribution for the critical current $I_c$, as resulting from the inclusion of stochastic effects, we discretize time and numerically simulate many trials of a switching current measurement in which $I_\text{bias}(t) = \dot I_\text{bias}t$ with a constant ramping speed $\dot I_\text{bias}$. At $t=0$, the washboard potential is initialized in accordance with the $\Phi$-dependent nature of the equilibrium many-body state discussed above. In the absence of parity transitions, at each time step $t\rightarrow t+dt$, the combined escape probability $(\Gamma_\text{TA}+\Gamma_\text{MQT})dt$ is compared with a random number to decide whether the junction switches to its resistive regime, in which case $I_c$ is recorded and the simulation loop is terminated. In most cases, the transition is dominated by thermal-activation at high temperatures and MQT at low temperatures, with a crossover temperature at which both are significant. To include the effects of parity transitions, $\Gamma_\text{P}dt$ is used to decide whether the washboard potential undergoes a transition between the two parity states as the bias current is swept. This has the effect of modifying the critical current distribution, in general shifting it to lower currents.

Our treatment of $p_x+ip_y$ superconductor devices is not based on a particular physical system or geometry. However, based on the material geometry and motivated by measurements we have made on the different but related S-TI-S lateral junction system implemented with Nb as the superconductor and Bi$_2$Se$_3$ as the topological insulator, we can expect Josephson junction critical currents of order $1\mu A$, junction resistances of $10-100\Omega$, and lateral junction capacitances of $0.1-10fF$. We can also estimate the size of the critical current carried by Majorana states in the junction, which sets the scale for the splitting of the parity state, a quantity measurable in the S-TI-S system as the lifting of odd-numbered nodes in the modulation of critical current with applied magnetic field. Using these parameters, we can calculate the expected critical current distribution for transitions dominated by thermal activation (TA), macroscopic quantum tunneling (MQT), or with both significant near the crossover between TA and MQT. We can also include the effects of Majorana pair parity transitions that occur during the measurement. 

\begin{figure}
    \centering
    \includegraphics[width=.48\textwidth]{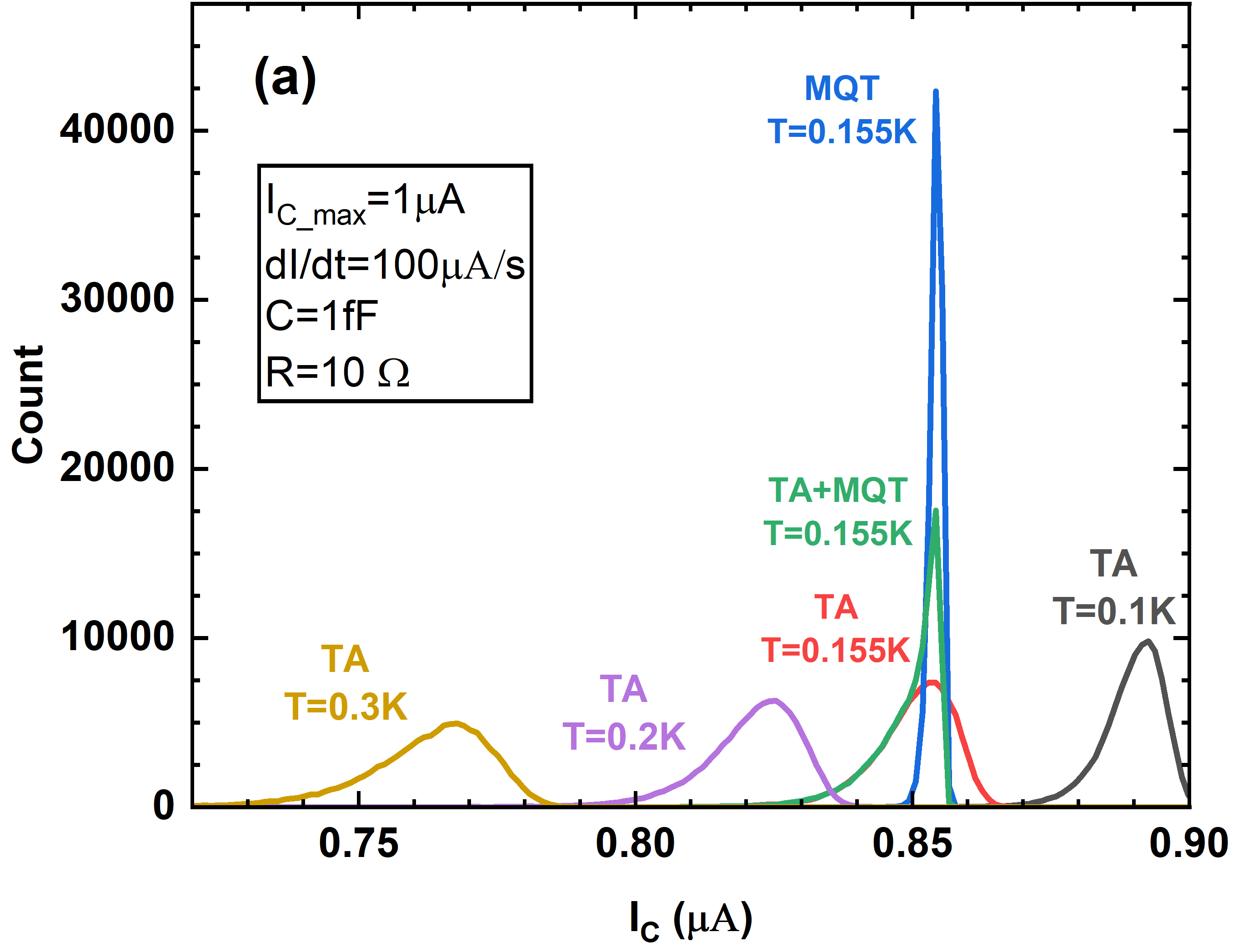}
    \includegraphics[width=.48\textwidth]{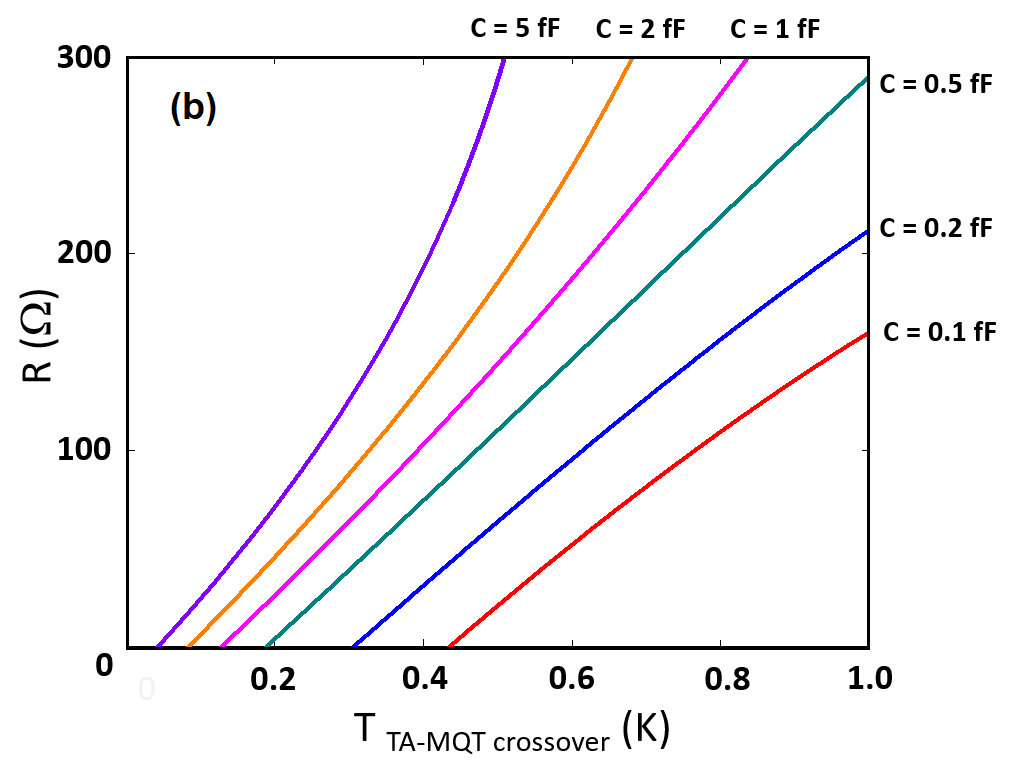}
    \caption{ Calibration of different regimes in regular junctions which do not have parity effects (a) Simulations of the critical current distributions for TA and MQT  at different temperatures. The temperature T=$155mK$ is the crossover temperature at which the TA and MQT distributions overlap. (b) The TA-MQT crossover temperature as a function of resistance for different values of the junction capacitance.}
    \label{fig:crossover}
\end{figure}

As a means of calibrating the associated parameter regimes, Figure \ref{fig:crossover}a plots simulated critical current distributions for C=$1fF$, R=$10\Omega$, deterministic critical current $I_{C\_max}=1\mu A$, ramp rate at $100\mu A/s$ based on the TA or MQT rate. Note that critical current distribution resulting from the inclusion of stochastic effects can differ significantly from the deterministic critical currents ($I_{C\_max}$). By comparing the plots at T= $100mK, 155mK, 200mK$, and $300 mK$, we show that the crossover from MQT-dominated to TA-dominated transition from the Josephson washboard well as the temperature increases. The critical current distribution generated by MQT does not change with temperature, while the distribution generated by TA is sensitive to temperature changes. We find at crossover temperature  T=$155mK$ the critical current distribution generated by MQT and TA overlaps, where both mechanisms contribute. As also shown in the same figure, the critical current distributions by MQT and TA have quite different shape, where the MQT one is much sharper than the TA ones. The distribution at crossover temperature has a shape from the mix of both mechanisms. Hence by observing the shape of critical current distribution of a Josephson junction can tell either TA or MQT mechanism is dominated at certain temperature. Figure \ref{fig:crossover}b shows the expected TA-MQT crossover temperature as a function of junction resistance for different values of the junction capacitance, obtained at temperature when TA and MQT critical current distributions overlap. Lower capacitance, corresponding to a smaller mass in the Josephson washboard potential enhances the MQT rate and raises the crossover temperature; lower junction resistance resulting in damping of the Josephson phase dynamics suppresses MQT and lower the crossover temperature. 

\begin{figure}
    \centering
    \includegraphics[width=.48\textwidth]{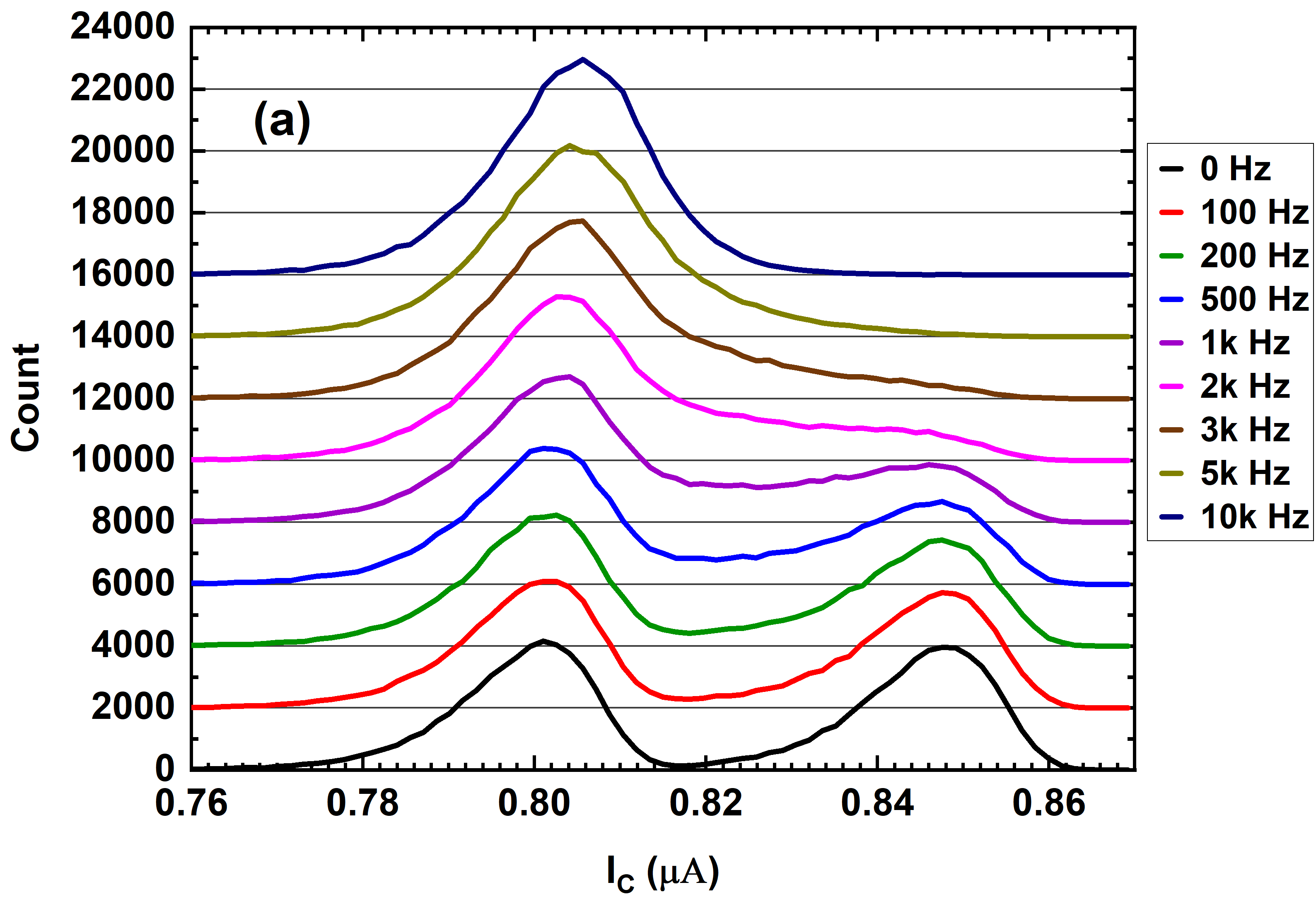}
    \includegraphics[width=.48\textwidth]{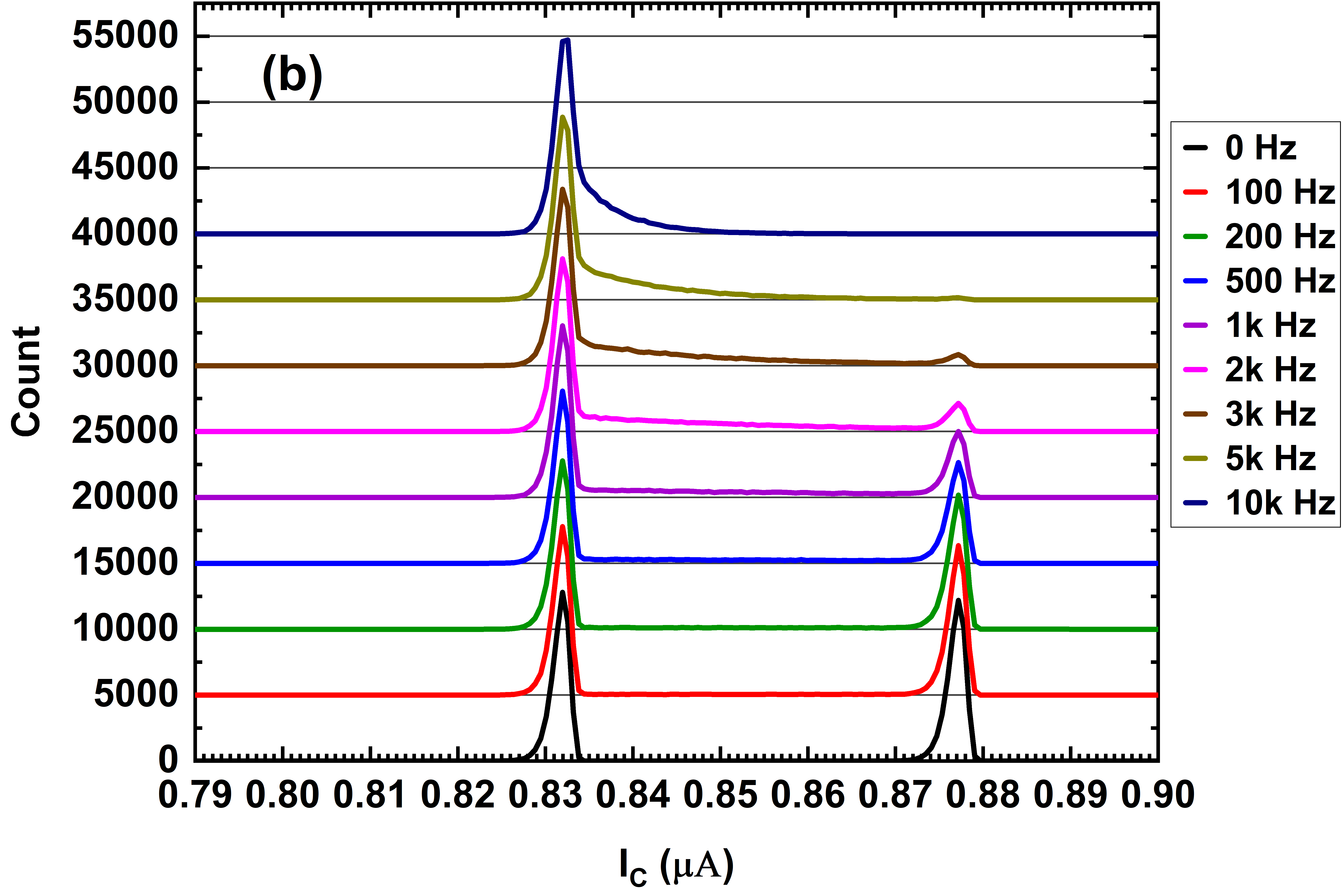}
    \caption{Evolution of critical current distribution and parity-based bimodal peaks for a range of parameters. Plots show distributions with $I_{C\_max}=1\mu A$ and different parity transition rates: (a) T=$200mK$, TA-dominated regime, (b) T=$100mK$, MQT-dominated regime}
    \label{fig:distribution}
\end{figure}

In Figure \ref{fig:distribution}, assuming the same parameters as the previous paragraph, we show the effect of parity fluctuations that induce premature transitions from the washboard well. In Figure \ref{fig:distribution}a, the critical current distribution is simulated at T=$200mK$ above the TA-MQT crossover temperature, such that the system has a TA-dominated transition. In the presence of MBS parity fluctuations, the distribution is bimodal. With critical current $I_C$ splitting equal to the difference in the critical currents of the two MBS states (the splitting is assumed $5\%$ of $I_C$ in the simulation). During the bias current sweep in the simulation, we allow the MBS parity states to fluctuate at a fixed average rate in time and assume both parity states appear at equal chances. Hence at certain time point the Josephson junction will have different TA rates based on the different split $I_C$ from the MBS parity state. For low parity transition rate, the double peaks $I_C$ distribution reflects that the system has the random initial parity state that persists throughout the bias current sweep. As the parity transition rate increases, parity has chance to switches into the lower parity state during bias current sweep, and this induces immediate transitions from the washboard well at currents intermediate between the two parity states. We show the evolution of the shapes of the distributions as function of the parity rate. The significance of this dependence is that it shows direct evidence for the parity degree of freedom generated by the Majorana bound states and enables a way to determine the parity transition rate from experiments. Similarly we plot the critical current distribution in MQT regime as shown in Figure \ref{fig:distribution}b for T=$100mK$. In the MQT regime, the double peaks feature becomes much sharper than the TA regime, and again the shape of the distribution involves as the parity transition rate changes.

We note that this qualitative behavior can be observed for a large range of parameters.  For example, for the system in consideration here, it is possible that the critical currents achievable in a proximity coupled $p_x+ip_y$ superconductor junction could be very small, perhaps in the nanoampere range which would be measurable only at ultra-low temperatures. This range is consistent with the scale derived from the tight-binding model in the previous subsection. However, in our simulations we see that the bimodal signatures for  Majorana parity states and parity fluctuation effects do persist and thus offer a robust approach to probe Majorana physics in topological systems.

\begin{figure}
    \centering
    \includegraphics[width=.48\textwidth]{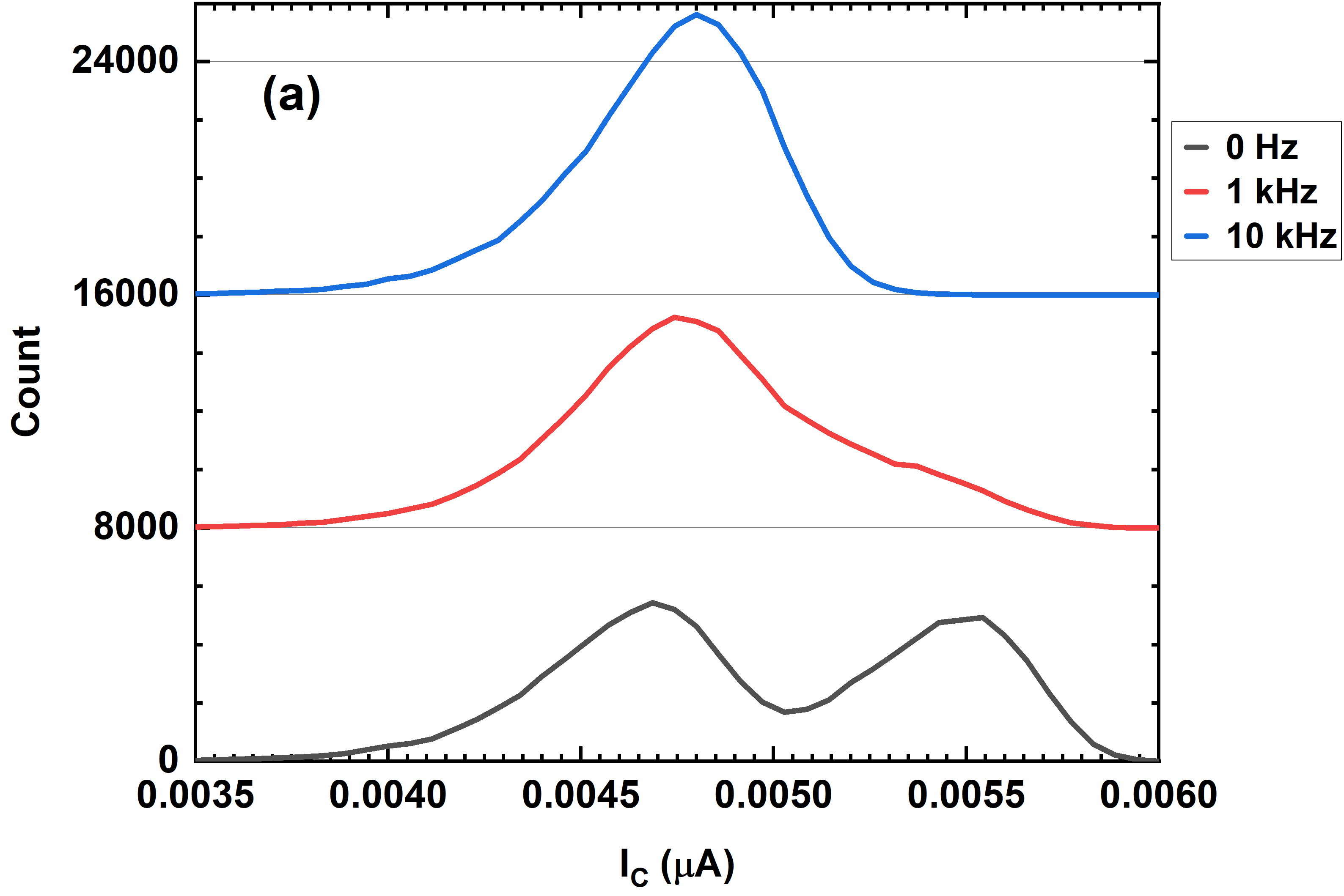}
    \includegraphics[width=.48\textwidth]{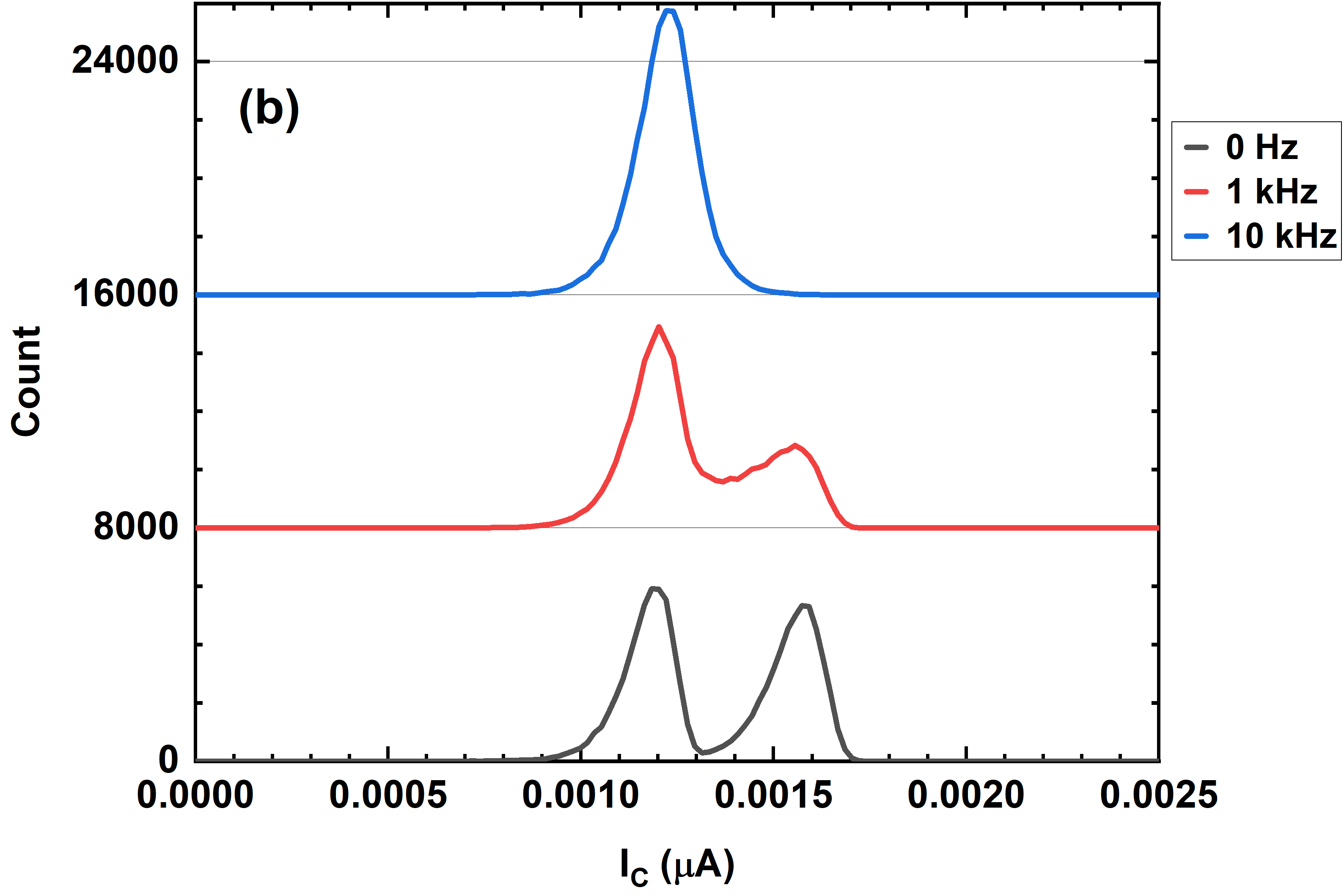}
    \caption{Evolution of critical current distribution and parity-based bimodal peaks for a second range of parameters. Plots show distributions with $I_{C\_max}=10 nA$ and different parity transition rates: (a) junction switches through TA at T=$10mK$ (b) junction switches through MQT}
    \label{fig:distribution_small}
\end{figure}

To demonstrate this persistence, we show in Figure \ref{fig:distribution_small}, simulations of the critical current distributions for these extreme conditions consistent with small critical current and parity splittings, choosing a junction with a critical current 100 x smaller, $I_{C\_max}=10nA$, and a resistance 10 x larger, R=$100\Omega$. The capacitance remains the same, C=$1fF$, consistent with a lateral Josephson junction between proximity-coupled superconducting islands. Last we assume the MBS parity states cause $10\%$ splitting of critical current. We show the distribution that would arise for dominance of thermal activation processes at $T=10mK$ and for MQT for parity transition rates of 0Hz, 1kHz, and 10kHz.  The qualitative nature of the distributions persist and provide a path to measure the parity lifetime.    

We note that there could be additional considerations in systems with low critical currents measured at ultra-low temperatures. In this regime, we expect that the equiprobable mixed initial state might no longer be prepared by thermal equilibration. In principle, the mixed initial state could be prepared by induced parity processes before each trial. These processes could involve controlled quasiparticle current injection, or tunneling from a voltage-tuned quantum dot into a localized MBS which could in turn be moved into the tunneling regime via flux manipulation. 

While the RCSJ treatment is general to parity-depended switching current derivations, the specific forms of the critical current and other parameters in the model directly employed the extended junction numerical data. The exact experimental parameters naturally depend on specific experimental geometries. Our results here are proof-of-principle that these bimodal current distributions are a consequence of parity and are expected to be robust within reasonable parameter ranges. 

\section{Trijunction and braiding}\label{trij}
A crucial aspect of Majorana physics is the ability for MBS to be braided via exchange. This leads to non-Abelian rotations in the basis of two-qubit parity states, an operation that is fundamental to the design of logical gates in any quantum computation platform \cite{Nayak}. These processes have been extensively analyzed in the T or Y junctions in nanowire systems \cite{FuKane, Alicea_2012}. While manipulating the position of Majorana states in nanowires require gating to move the interface between topoogical and trivial regimes, the MBS in the lateral Josephson junctions considered here are bound to the Josephson vortices and can be easily moved by adjusting the phases in the junctions by currents and voltages.  A possible trijunction geometry is shown in Fig. \ref{2DJ}, consisting of uniform channels between three superconducting islands.  Magnetic field applied to the device creates a gradient in the phase across the junctions which results in the entry of Josephson vortices and MBS bound to them into the junction. In the absence of external biases, the Josephson vortices/MBS are symmetrically located in the junctions, spaced by one flux quantum. By adjusting the relative phases between the islands, the bound states may be moved along the junction. We note that Ref.~\onlinecite{choi2018} proposes a probe of non-Abelian statistics in a closely related system.

The change in the phase and subsequent movement of the MBS can be achieved in two ways: (1) By applying a current that exceeds the critical current, the junction can be driven into the normal state in which there is a finite voltage across the junction.  This causes the phase to wind in time according to the Josephson relation, $V=(\hbar/2e)\frac{\partial\phi}{\partial t}$ and moves the Josephson vortices and MBS linearly in time. By applying appropriate voltage pulse sequences, we can manipulate the MBS positions. (2) By applying a dc supercurrent through the junction to create static phase difference.  By shunting the junction with a superconducting inductor, any phase difference can be generated, not limited by the critical current of the junction, and the position of the Josephson vortices and MBS can be continuously controlled. Each of these schemes has advantages.  The first approach allows measurements of the junction critical currents and can utilize the established technology of RSFQ (Rapid Single Flux Quantum) pulse technology common in digital Josephson junction circuits). However, because the junctions are driven into a finite voltage state, quasiparticles may be generated that could enhance parity fluctuations via quasiparticle poisoning.  The second approach avoids this problem because the junction are never driven into a finite voltage state, avoiding quasiparticle production.  However, this is at the expense of shorting the entire circuit with superconductors, making it more difficult to track the location of the vortices via transport measurements and instead requiring phase-sensitive SQUID measurements or scanning probe microscopies.    

\begin{figure}
    \centering
    \includegraphics[width=.48\textwidth]{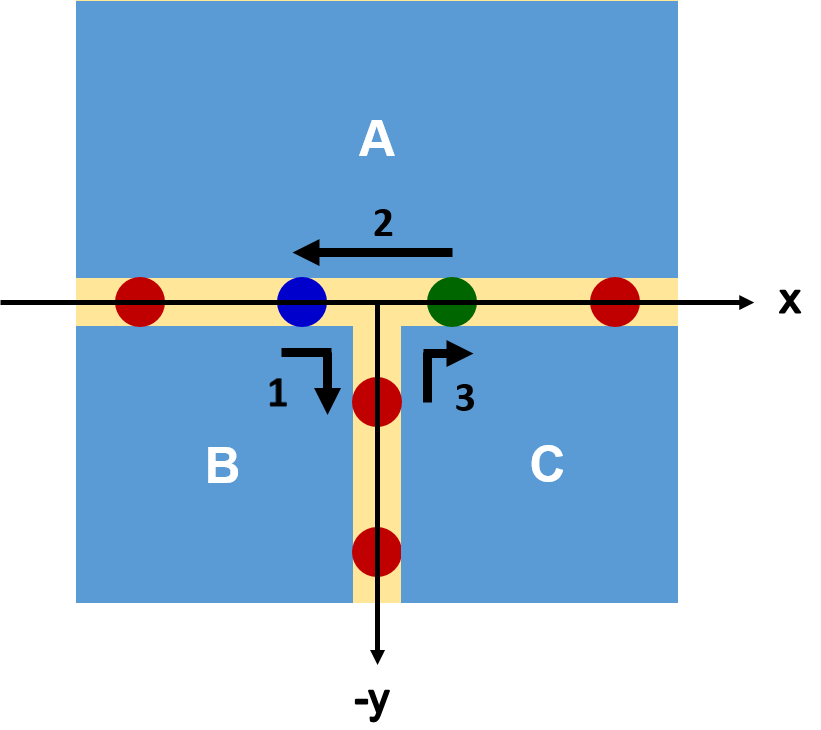}
    \caption{Three-step operation to affect a braiding exchange of the Majorana states shown in blue and green via a sequence of single-flux quantum voltage pulses that move the host Josephson vortices.}
    \label{fig:braid}
\end{figure}

Here, we focus on the voltage pulse scheme.  We can perform an exchange as shown in Figure \ref{fig:braid}.  As defined also in Figure \ref{2DJ}b, let the horizontal junctions be along the x-axis, the vertical junction along the negative y-axis and the point of conjunction be the origin. We can parameterize the gauge-invariant phase difference across the junctions in the presence of an applied vertical magnetic field [B] as 
$\varphi_{BA}(x)=\phi_{0,BA}- Fx$, $\varphi_{AC}(x)=\phi_{0,AC}+Fx$ and  
$\varphi_{CB}(y)=\phi_{0,CB}-Fy$. Here, $\phi_{0,IJ}$ is the phase difference at the origin for the junction between islands I and J, and $F=\frac{2\pi}{W}\frac{\Phi_{IJ}}{\Phi_0}$ is the magnetic flux per length threading each junction.

Now, to provide a specific scheme for enabling such an exchange, let us say a voltage pulse of amplitude $V$ is applied for a time $\Delta t$ on the island $B$ relative to islands $A$ and $C$ islands. This causes the phases to evolve in time in the $AB$ and $BC$ junctions according to the Josephson relation, with $\phi_{0,BA}$ being modified to $\phi_{0,BA}+\frac{2e}{\hbar}V\Delta t$ and $\phi_{0,CB}$ being modified to $\phi_{0,CB}-\frac{2e}{\hbar}V\Delta t$. This results in the position of all of the Josephson vortices and the Majorana states bound to them in the $AB$ and $BC$ junctions to move by a distance $+\frac{2e}{\hbar}\frac{V\Delta t}{F}$.  In particular, if $V\Delta t = \Phi_0$, each Majorana/Josephson vortex moves by the vortex spacing. This can be achieved experimentally by applying pulses of magnitude $V \approx 1\mu V$ and duration $\Delta t \approx 1ns$. Repeating this operation by applying a voltage pulse to island {A} and then island {C}, we can achieve the exchange of the two Majoranas as in Figure \ref{fig:braid}.  Details of this process are elucidated in Ref.~\onlinecite{Hegde2020}. It must be stressed here that application of voltage pulses in a practical setting could result in quasiparticle poisoning events, leading to breakdown of parity conservation and hence, the preservation of the MBS. This can be avoided with the current pulse scheme.

Turning to gate operations, the simplest one that can be accomplished by exchange braiding can be constructed using four Majoranas, for instance arranged along the $x$ direction as illustrated in Figure \ref{fig:braid}b. We can label the four Majoranas as $\gamma_i$ at position $s_i$, with $i=a,b,c,d$. Assuming the total fermion parity to be conserved, \emph{i.e.}, $(-)\gamma_a\gamma_b\gamma_c\gamma_d=-1$, we have two states forming a qubit in the Hilbert space: $|1_{ab}0_{cd}\rangle$ and $|0_{ab}1_{cd}\rangle$, where we use the notation $|0_{ij}\rangle$ or $|1_{ij}\rangle$ to represent a state of 0 or 1 fermion created by $(\gamma_i-i\gamma_j)/2$. Upon the application of a voltage pulse, the system returns to the same Hilbert space after the braiding operation between $\gamma_b$ and $\gamma_c$. This adiabatic evolution generally applies a quantum gate $U$ to the system: $U|\psi_i\rangle=|\psi_f\rangle$. Instead of using the Schr\"{o}dinger picture, we can equivalently use a Heisenberg picture. The braiding operation that swaps $\gamma_\beta$ and $\gamma_3$ must satisfy:
\begin{equation}
    U^{\dagger}\gamma_{a,d}U=\gamma_{a,d},\quad U^{\dagger}\gamma_b U=\eta \gamma_c,\quad U^{\dagger}\gamma_c U=\theta \gamma_b,
\label{eq:transform}
\end{equation}
where $\eta$ and $\theta$ must take value of $\pm 1$ so that the Majorana operators are real. Due to the conservation of fermion parity we have $\eta\theta=-1$. Up to a gauge choice, we can fix $\eta=1$ and $\theta=-1$. Now from the transformation law Eq.~\eqref{eq:transform}, the transformation matrix $U$ has to take the form $U=\cos\phi+\sin\phi \gamma_b\gamma_c$ and we can solve for $\phi=\frac{\pi}{4}$. We can see that this indeed is a nontrivial single qubit gate:
\begin{equation}
\begin{split}
      U|1_{ab}0_{cd}\rangle&=\frac{1}{\sqrt{2}}(|1_{ab}0_{cd}\rangle-i|0_{ab}1_{cd}\rangle)\\
    U|0_{ab}1_{cd}\rangle&=\frac{1}{\sqrt{2}}(-i|1_{ab}0_{cd}\rangle+|0_{ab}1_{cd}\rangle),
\end{split}
\end{equation}
which is a single qubit gate that acts on the Majorana qubit. Demonstrating such non-Abelian operations would require, in addition to exchange, such as via the pulse sequence discussed here, a viable parity qubit read-out scheme. 

\section{Discussion}\label{disc} In summary, we have presented a detailed analysis of extended 2D topological superconducting junctions, the associated spectrum of low-lying states, and the appearance of the sought-after MBS on application of flux through the junction. Amidst the tangle of states, the localized MBS are indeed prevalent, and their evolution as a function of applied flux and global phase could potentially be detected through various means, such as quantum dot spectroscopy. The associated zero-energy electronic state and parity-dependent many-body ground state give rise to palpable signatures in Josephson current Fraunhofer patterns and bimodal switching current distributions. Scaling up to have networks of extended Josephson junctions provides an entire platform for applications to Majorana-based quantum computation. Here, we show one possible scheme for MBS exchange in trijunction geometries that give rise to non-Abelian rotation. 

The work presented here opens up several theoretical issues to be resolved, as has been done extensively in the nanowire setting. To mention a few, in contrast to nanowires, where induced superconductivity gives rise to a distinct gap (in the absence of strong disorder), the extended junction carries with it a whole spectrum of low-lying dispersive modes. Thus, care is required in isolating MBS features from the rest of this spectrum, and stringent diagnostics and bounds are yet to be developed. We proposed quantum dots to detect the presence and evolution of MBS;  further work would model the nature of the coupling between the dot and junction degrees of freedom as well as the dot itself. With regards to stochastic processes in the RCSJ model, we introduced three rates phenomenologically; in principle, future work can derive these rates from within the model itself. Turning to the trijunction scheme, the time-dependent pulse application and motion of MBS would require a whole non-equilibrium treatment for a full-fledged analysis. 

On the experimental front, there has been a keen hunt for materials that exhibit intrinsic unconventional p-wave paired superconductivity over the last few decades. The difficulty in detecting p-wave paired states stems from their similarity to s-wave states in the thermodynamic limit\cite{Mackenzie,Flouquet2005OnTH}. Several promising candidates continue to be investigated, chief among them being Sr$_2$RuO$_4$\cite{Ishida,Luke,Rice}. While there were early indications that the material most likely exhibited triplet pairing, more recent NMR experiments offer evidence to the contrary \cite{Kallin_2012,Pustogow2019}. There has also been conflicting evidence regarding chirality and the nature of time reversal symmetry breaking in Sr$_2$RuO$_4$\cite{Luke, Ishida,VanHarlingen,Kirtley,Liu_2010,Nelson, Pustogow2019,Abergel2019}. Other candidates are also being actively pursued, such as iron-based superconductors\cite{Wang,Zhang,Zhang2019,Wang2021,Talantsev2019}, bilayer BiH\cite{Yang} and monolayer graphene\cite{DiBernardo}. While materials that possess intrinsic $p$-wave superconducting pairing are hard to find in nature, hybrid structures can provide alternate realizations. In the case of nanowires, effective p-wave pairing is often achieved through the proximity effect as well as Rashba/Dresselhaus spin-orbit coupling. In the case of lateral junctions, as presented here, the $p_x+ip_y$ element assumes time-reversal symmetry breaking. Such a scenario has been introduced in Refs.~\onlinecite{choi2018,choi2019} and also proposed~\cite{Qi:2010,Chung:2011,Wang:2015,Lian} in certain quantum anomalous Hall platforms~\cite{Chang:2013,He:2017,Kayyalha:2020,Shen:2020}. These models are strictly 2D, and our results are applicable in locating the ``missing'' Majorana delocalized around the system's perimeter.

The original setting where extended junction MBS were conceived \cite{FuKane,FuPotter} offer another highly promising set of platforms. These platforms entail proximity-induced superconductivity on the surface of a 3D topological insulator (TI), forming S-TI-S junctions. Although the low energy electronic states on the Fermi surface have an effective $p_x+ip_y$ pairing, the topology of the system is determined by the full ground state wavefunction and this scenario does not resemble a chiral topological superconductor due to the preserving of time-reversal symmetry. To apply our model to these systems, a time-reversal breaking insulating gap has to be open around the superconducting regime to produce the chiral Majorana fermions around the island. Even then, only some characteristic features presented here apply due to the presence of two interfaces in the direction transverse to the plane. Specifically, while we do not expect the delocalized Majorana edge mode due to partnering at the two interfaces, 
we remark that the theoretical studies performed here in fact stemmed from considerations of the S-TI-S junction geometries realized by the experimental group of Van Harlingen. Their initial work showed evidence of a topological phase transition in the Josephson current running through such junctions\cite{Stehno}. Our aforementioned theoretical work in the Introduction on a complete architecture for manipulating MBS in lateral junctions\cite{Hegde2020} was conducted in collaboration with this group based on their current capabilities. 

In conclusion, currently there is fertile theoretical and experimental ground for exploring MBS in lateral junction geometries. Here, we have provided directions for detecting the evolution of MBS with applied flux, parity signatures in bimodal distributions, and a scheme for performing MBS exchange. Demonstrating such exchanges, related non-Abelian rotations, and parity-controlling qubit gate operations is the ultimate goal. Identifying the optimal experimental setting for bimodal signatures,achieving parity readout, be it detection with quantum dot devices, transmon circuits, or other possibilities, and exchanges would each constitute a challenge and if met with success,a tremendous leap.
\begin{acknowledgments}
We are grateful to detailed discussions with Suraj Hegde over the course of this project and to Alex Levchenko for  illuminating conversations. We acknowledge the support of the National Science Foundation through  grant DMR-2004825 (NA, SV, DVH) and the Quantum Leap Challenge Institute for Hybrid Quantum Architectures and Networks grant OMA-2016136)(VS, GY). XQS acknowledges  support from the Gordon and Betty Moore Foundation's EPiQS Initiative through grant GBMF8691.
\end{acknowledgments}

\bibliography{references}
\end{document}